\documentclass[12pt]{article}
\usepackage{graphicx}
\usepackage[latin1]{inputenc}
\usepackage{amsfonts}
\usepackage{epsfig}
\usepackage{amsmath}
\usepackage{latexsym}
\usepackage{amssymb}
\usepackage{graphicx}

\begin{document}

\title{\normalsize\bf
Viscosity and Structure Configuration Properties \\ of Equilibrium and Supercooled Liquid Cobalt}
\author{\normalsize R. M. Khusnutdinoff$^1$\footnote{Corresponding author.
E-mail: {\tt khrm@mail.ru}}, A. V. Mokshin$^1$\footnote{Corresponding author.
E-mail: {\tt anatolii.mokshin@mail.ru}}, A. L. Beltyukov$^2$, N. V. Olyanina$^2$\\
\normalsize\emph{$^1$Kazan (Volga region) Federal University, 420008 Kazan, Russia}\\
\normalsize\emph{$^2$Physical-Technical Institute, Ural Brunch of Russian Academy of Sciences,} \\
\normalsize\emph{426000 Izhevsk, Russia}}

\date{\normalsize (Received on 1 February 2018)}
\maketitle

\begin{abstract}
The shear viscosity of liquid cobalt at the pressure $p=1.5$~bar and at the temperatures
corresponding to equilibrium liquid and supercooled liquid states is measured experimentally and evaluated by
means of molecular dynamics simulations. Further, the
shear viscosity is also calculated within the microscopic
theoretical model. Comparison of our experimental, simulation
and theoretical results with other available data allows one to examine
the issue about the correct temperature dependence of the shear
viscosity of liquid cobalt. It is found a strong correlation between
the viscosity and the configuration entropy of liquid cobalt
over the considered temperature range, which can be taken into account by the
Rosenfeld's model.

%\begin{keyword}
%Liquid cobalt \sep molecular dynamics \sep entropy
%\sep viscosity \sep scaling law
%\PACS
%61.20.Ja \sep 61.25.Em \sep 68.35.Rh
%\end{keyword}
\end{abstract}

%61.20.Ja   Computer simulation of liquid structure
%61.25.Em   Molecular liquids
%68.35.Rh   Phase transitions and critical phenomena

%\date{\today}

\section*{\normalsize Introduction}
Cobalt inferior to iron, aluminium and copper in the degree of use
in the ordinary metallurgical industry. Nevertheless, due to its
physical properties, cobalt is applied in the aerospace
industry. In particular, cobalt-based alloys stand out by the following properties as high strength, corrosion resistance and hardness extended over a wide temperature range. Similar to a simple single-component system, pure cobalt
is not inclined to generate an amorphous phases. Cobalt melt crystallizes
at low and moderate levels of the supercooling, and at values of
the viscosity $\eta$ much smaller than $\eta\simeq
10^{12}$~Pa$\cdot$s. Thus, the deep supercooling levels with the
temperatures lower than the glass transition temperature $T_g$ are
not attainable for the case of cobalt. Nevertheless, as follows
from recent results~\cite{Book_AMA,Book_BMG}, the inclusion of
cobalt in the composition of metallic melts can
enhancement of their glass-forming abilities.

The viscosity is one of the main characteristic of a system, which
defines its crystallization and glass-forming ability \cite{Gaskell1989,Khusnutdinoff2013,Khusnutdinoff2017}. In
particular, the viscosity determines directly the
crystal nucleation and growth rates, as well as the ability to
generate and to retain a disordered state~\cite{Fokin_2006}. For the
case of liquid cobalt, there is an ambiguous situation with regard
to temperature dependence of the viscosity~$\eta(T)$. The known
experimental data for the viscosity $\eta(T)$ of liquid cobalt~
\cite{Iida/Guthrie,Assael2012,HandBook} have a difference in
values about 30\% and even more. To our knowledge,
there are no experimental data on the viscosity of supercooled cobalt. Therefore, one of the
purposes of the given study is  to refine values of the shear
viscosity of cobalt for the equilibrium liquid phase
(at the temperatures higher the melting
temperature~$T_m=1768$~K, where $\eta(T_m)=5.4$~mPa$\cdot$s
\cite{Iida/Guthrie}) and to evaluate the viscosity of the
supercooling liquid cobalt (with the temperatures $T<T_m$). Thus,
we present experimental and molecular
dynamics simulation results of the shear viscosity of liquid cobalt
for the temperature range $T=[1400;\;2000]$~K and at the pressure
$p=1.5$~bar. Furthermore, the temperature dependence $\eta(T)$ is
also calculated within the microscopic theoretical model where
the parameters are computed on the basis of the configuration
data of the system.

As follows from statistical mechanics
\cite{Pitaevskii/Lifshitz,Evans/Morris}, the viscosity of a
(dense, condensed) system is defined by the interparticle
interaction energy and by the structural characteristics. This is
clearly seen from the microscopic expressions for the viscosity
suggested by Born and Green \cite{Born/Green} and
by Irving and Kirkwood \cite{Irving/Kirkwood} and applied usually
to compute the viscosity on the basis of molecular dynamics
simulations data. Further, pronounced correlation effects between
the viscosity and the configuration entropy, which are accounted
for by the Rosenfeld's and Dzugutov's relations, were
established for a various metallic
melts~\cite{Dzugutov_1996,Li_2005,Samanta_2005,Ali_2006,Fomin_2012,Gosh_2013,Cao_2014,Pasturel_2015}
and a lot of molecular liquids under high pressure~\cite{Abramson_2007,Abramson_2008,Abramson_2009,Abramson_2011,Abramson_2014}.
Therefore, the correlation between the viscosity and the structure
configuration properties of liquid cobalt over whole the
considered temperature range is also verified in the given study.

\section*{\normalsize Experimental Details \label{expt}}

The viscosity  of liquid cobalt for the temperatures over the
range from $T=1506$ to $1969$~K has been experimentally measured
by the method of torsional vibrations of a cylindrical crucible
with the melt~\cite{Beltyukov2008,Khusnutdinoff2016}. The mass fraction of cobalt in
the samples was no less than 99.98$\%$, while the samples could
also contain the following impurities: less than 0.003$\%$ of
Fe, less than 0.005$\%$ of Ni and C, less than  0.001$\%$ of Si,
Cu, Mg, Zn and Al, and less than  0.001$\%$ of O. The cylindrical
crucibles with the internal diameter $\sim17$~mm and  the
height $\sim42$~mm  were made from $\rm{Al_2O_3}$. The temperature
dependence of the viscosity was measured by heating of the
samples from the melting temperature $T_m$ up to the temperature
$T=1973$~K. Then, the samples were cooled until they crystallized.
The temperature step of the heating/cooling procedure was $\Delta T=20\pm5$~K, and the melt samples were
equilibrated at each the temperature over twenty minutes.
Experimental value of the temperature was determined with the precision $\pm
5$~K by means of the tungsten-rhenium thermocouple calibrated for
the melting points of pure Al, Cu, Ni and Fe.

The kinematic viscosity $\nu$ was evaluated by means of numerical
solution of the motion equation of the cylindrical
crucible~\cite{Beltyukov2008,Khusnutdinoff2016}:
\begin{equation}
\textrm{Re}[L(\nu)]+\frac{\delta}{2\pi}\textrm{Im}[L(\nu)]-2I\bigg(\frac{\delta}{\tau}-\frac{\delta_0}{\tau_0}\bigg)=0.
\end{equation}
where $I$ is the moment of inertia; $\delta$ and $\tau$ are the
attenuation decrement and the oscillation period of the system
with the melt, whereas  $\delta_0$ and $\tau_0$  are the same
characteristics for the empty crucible, respectively. Finally,
$\textrm{Re}[L]$ and $\textrm{Im}[L]$ are the real and imaginary
parts of the friction function $L$, which is related to the
kinematic viscosity $\nu$ (see Ref.~\cite{Beltyukov2008}, for
details).

The average linear coefficient of the thermal expansion for the
crucible material $\rm{Al_2O_3}$ over the temperature range from
$273$~K to $2073$~K  takes the value $9.0\cdot10^{-6}$
deg.$^{-1}$ (Ref.~\cite{RefBook}), and the possible impact of the thermal
expansion on the measured values of the viscosity must be also taken
into account. The height of the melt within the crucible is determined as
\begin{equation}
H=\frac{M}{\pi R^2\rho_{\mu}}.
\end{equation}
Here, $M$ is the mass of the sample, and $\rho_{\mu}$ is the mass
density of the melt, which was defined from the relation
\begin{equation}
\rho_{\mu}(T)=6172.152-0.936T.
\end{equation}
suggested in Ref.~\cite{Assael2012}.

\section*{\normalsize Details of Simulation and Numerical Calculation \label{comp}}
Molecular dynamics simulations of liquid cobalt were carried out for the isothermal-isobaric (NpT)-ensemble for the temperatures from the range $T=[1400; 2000]~$K and at the pressure $p=1.5$~bar. The system was consisted of $N=4000$ atoms located in a cubic cell with periodic boundary conditions. Interaction between atoms was carried out using the EAM-potential \cite{Passianot_1992,Belashchenko2013}.
The equations of motion for atoms were integrated using the velocity Verlet algorithm with the time step $\tau=1.0$~fs \cite{Gonzalez20001}. To bring the system into a state of thermodynamic equilibrium and to calculate the temporal and spectral characteristics of the system, it was realized the dynamics with $100\;000$ and $2000\;000$ time steps, respectively.

The shear viscosity can be determined in the framework of the Green-Kubo approach
\cite{Hansen/McDonald} through the autocorrelation functions of the stress tensor, $\sigma_{\alpha,\beta}$.
The non-diagonal components of the stress tensor are given by \cite{Boon/Yip}
\begin{eqnarray}
\sigma_{\alpha,\beta}=\frac{1}{V}\bigg(\sum_{i=1}^N m\vartheta_{i\alpha}\vartheta_{i\beta}-
\sum_{i=1}^{N-1}\sum_{j=i+1}^{N}r_{ij\alpha}\frac{\partial U(r_{ij})}{\partial r_{ij\beta}} \bigg), \label{Eq_Vars}
\end{eqnarray}
where $\vec{r}$, $\vec{\vartheta}$ are the position and the velocity of a particle, $i, j$ are the
numbers of particles, $\vec{r}_{ij}=\vec{r}_{i}-\vec{r}_{j}$, $U(r_{ij})$ is the particle interaction
potential, $\alpha, \beta$ are the indices of the components of the corresponding vectors; $m$ is the particle mass and $V$ is the volume of the system.
Then, the shear viscosity can be calculated by the formula
\begin{equation}
\eta=\frac{V}{k_BT}\int_0^{\infty}\langle \sigma_{\alpha,\beta}(t)\sigma_{\alpha,\beta}(0) \rangle dt, \label{eta_s}
\end{equation}
where $k_B$ is the Boltzmann constant.

\section*{\normalsize Theoretical Formalism}
Let us consider a system consisting of $N$ identical particles of mass $m$ enclosed in the volume $V$. We take the off-diagonal component of the stress tensor $\sigma_{\alpha,\beta}$, defined by Eq. (\ref{Eq_Vars}), as the initial dynamic variable. Then, we determine the time correlation function (TCF) of the stress tensor as follows \cite{Balucani_1994}
\begin{equation}
S(t)=\frac{\langle \sigma_{\alpha,\beta}(t)\sigma_{\alpha,\beta}(0) \rangle}{\langle |\sigma_{\alpha,\beta}(0)|^2 \rangle},
\end{equation}
and its spectral density as the next \cite{March_1990}:
\begin{equation}
\tilde{S}(\omega)=\frac{S_0}{2\pi}\textrm{Re}\int_{-\infty}^{\infty}e^{i\omega t}S(t)dt.
\end{equation}
Here, $S_{0}$ is the zeroth frequency moment of $\tilde{S}(\omega)$~(Ref.~\cite{Forster_1968}):
\begin{equation}
S_{0}=\langle |\sigma_{\alpha,\beta}(0)|^2 \rangle=\Bigg(\frac{k_BT}{V}\Bigg)^2+\frac{2\pi \rho}{15}\frac{k_BT}{V^2}\int_0^{\infty}r^4g(r)\bigg(\frac{4}{r}\frac{\partial \mathcal{U}}{\partial r}+\frac{\partial^2 \mathcal{U}}{\partial r^2}\bigg)dr,
\label{Eq_S0}
\end{equation}
$\rho$ is the number density, and $g(r)$ is the pair radial distribution function.

On the other hand, according to the formalism of the time correlation functions \cite{Hansen/McDonald, Mokshin_JETP}, the spectral density of TCF of the stress tensor $\tilde{S}(\omega)$ can be represented as infinite continuous fraction:
\begin{equation}
\tilde{S}(\omega)=\frac{1}{\pi}\textrm{Re}\left\{\displaystyle\frac{S_0}{i\omega+\displaystyle\frac{\Delta_1}{i\omega+\displaystyle\frac{\Delta_2}{i\omega+\ldots}}}\right\}.
\end{equation}
Here, $\Delta_n$, $n=1,2,3,... $ are the relaxation parameters, which are related with the frequency moments $S^{(2m)}$ of $\tilde{S}(\omega)$:
\begin{equation}
S^{(2j)}=\frac{\displaystyle\int \omega^{2j}\tilde{S}(\omega)d\omega }{\displaystyle \int \tilde{S}(\omega)d\omega }, ~~~~~~~ j=1,2,\ldots
\end{equation}
by means of the following expressions:
\begin{eqnarray}
\nonumber
 \Delta_{1}&=&S^{(2)},
 \nonumber\\
\Delta_{2}&=&\frac{S^{(4)}}{S^{(2)}}-S^{(2)},\nonumber\\
\Delta_{3}&=&\frac{S^{(6)}S^{(2)}-S^{(4)^{2}}}
{S^{(4)}S^{(2)}-S^{(2)^{3}}},
\ldots. \nonumber
\end{eqnarray}
The relaxation parameters can be determined numerically from molecular dynamics simulation data in accordance with the basic definitions \cite{Mokshin_JPCM}:
\begin{equation}
\Delta_n=\frac{\langle |A_n(0)|^2\rangle}{\langle |A_{n-1}(0)|^2\rangle}, ~~~ n=1,2\ldots,
\label{Delta_s}
\end{equation}
where
\begin{eqnarray}
A_0(t)&=&\sigma_{\alpha,\beta}(t), \nonumber\\
A_1(t)&=&\frac{\partial A_0(t)}{\partial t},  \nonumber\\
A_2(t)&=&\frac{\partial A_1(t)}{\partial t}+\Delta_1A_0(t), \nonumber\\
\ldots,   \nonumber\\
A_n(t)&=&\frac{\partial A_{n-1}(t)}{\partial t}+\Delta_{n-1}A_{n-2}(t). \label{Vars_s}
\end{eqnarray}

As shown in Refs. \cite{Mokshin_2003,Mokshin_2011}, alignment of the relaxation scales $\Delta_2^{-1/2}\approx\Delta_3^{-1/2}\approx\Delta_4^{-1/2}$ is observed for the transport processes in monatomic liquids. This allows one to obtain expression for the spectral density $\tilde{S}(\omega)$:
\begin{equation}
\tilde{S}(\omega)=\frac{1}{\pi}\frac{2\Delta_1\Delta_2\sqrt{4\Delta_2-\omega^2}}{\Delta_1^2(4\Delta_2-\omega^2)+\omega^2(2\Delta_2-\Delta_1)^2}.
\label{Eq_SoS}
\end{equation}
Then, in accordance with the Green-Kubo formula (\ref{eta_s}), we obtain the following expression for the shear viscosity:
\begin{equation}
\eta=\frac{VS_0}{\pi k_BT}\frac{\sqrt{\Delta_2}}{\Delta_1}.
\label{Eq_KinVisc}
\end{equation}

Taking into account Eq. (\ref{Eq_S0}) for $S_0$ and neglecting the kinetic contribution, one can see that  Eq. (\ref{Eq_KinVisc}) transforms into the Rice-Kirkwood approximated equation for  the viscosity at the melting temperature \cite{Rice/Kirkwood}:
\begin{equation}
\eta=\frac{2\pi m \rho^2}{15\zeta_f}\int_0^{\infty}r^4\bigg(\frac{4}{r}\frac{\partial \mathcal{U}}{\partial r}+\frac{\partial^2 \mathcal{U}}{\partial r^2}\bigg)g(r)dr,
\end{equation}
where the friction coefficient $\zeta_f$ is equal to
\begin{equation}
\zeta_f=\frac{\pi m \rho V\Delta_1}{\sqrt{\Delta_2}}.
\end{equation}
On the other hand, using the expression for the viscosity (7.32) from Ref. \cite{Iida/Guthrie/2015}
\begin{equation}
\eta\approx 4.38\omega_0 m \rho^2 g(r_m)r_m^5(1-r_0/r_m)
\label{Eq_ViscM}
\end{equation}
and the viscosity values obtained from formula (\ref{Eq_KinVisc}), one can estimate the characteristic frequency of atomic vibration $\omega_0$:
\begin{equation}
\omega_0=\frac{VS_0\sqrt{\Delta_2}}{4.38\pi k_BT\Delta_1m\rho^2g(r_m)r_m^5(1-r_0/r_m)}.
\label{Eq_FreQ}
\end{equation}
Here, $r_0$ is the minimum possible distance between neighboring atoms [where $g(r)$ starts to take nonzero values], $r_m$ is location of the main maximum of $g(r)$.

\section*{\normalsize Results}
%~~~~~~~~~~~~~~~~~~~~~~~~figure~~~~~~~~~~~~~~~~~~~~~~~~~~~~~~~~~~~~~~~~~~~~
\begin{figure*}
\begin{center}
\includegraphics[height=10cm, angle=0]{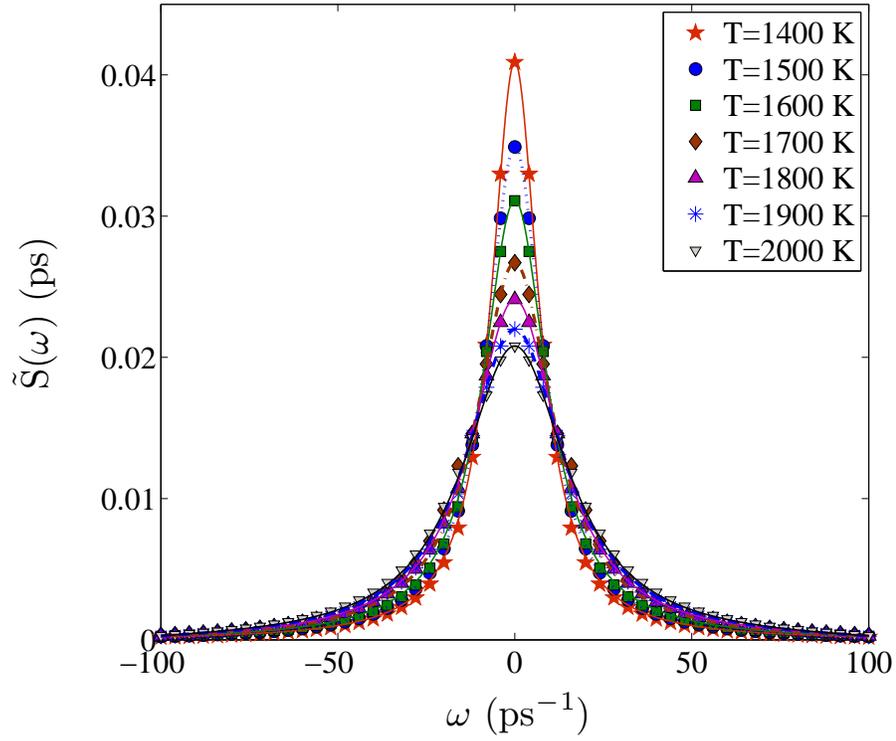}
\caption{(Color online) Spectral density of the stress tensor TCF of cobalt melt at different temperatures: markers represent simulation data; lines -- the results of theoretical calculations with the expression (\ref{Eq_SoS}).} \label{Fig_Stress}
\end{center}
\end{figure*}
%~~~~~~~~~~~~~~~~~~~~~~~~figure~~~~~~~~~~~~~~~~~~~~~~~~~~~~~~~~~~~~~~~~~~~

In Fig. \ref{Fig_Stress}, the spectral density $\tilde{S}(\omega)$ of liquid cobalt obtained molecular dynamics simulations results is presented and compared with theoretical results [equation (\ref{Eq_SoS})] at various temperatures. The relaxation parameters $\Delta_1$ and $\Delta_2$ were determined numerically from Eqs. (\ref{Delta_s}) and (\ref{Vars_s}). As seen from Fig. \ref{Fig_Stress}, theoretical curves reproduce correctly the spectra $\tilde{S}(\omega)$ for the considered temperature range.
%~~~~~~~~~~~~~~~~~~~~~~~~figure~~~~~~~~~~~~~~~~~~~~~~~~~~~~~~~~~~~~~~~~~~~
\begin{figure*}
\begin{center}
\includegraphics[height=11cm, angle=0]{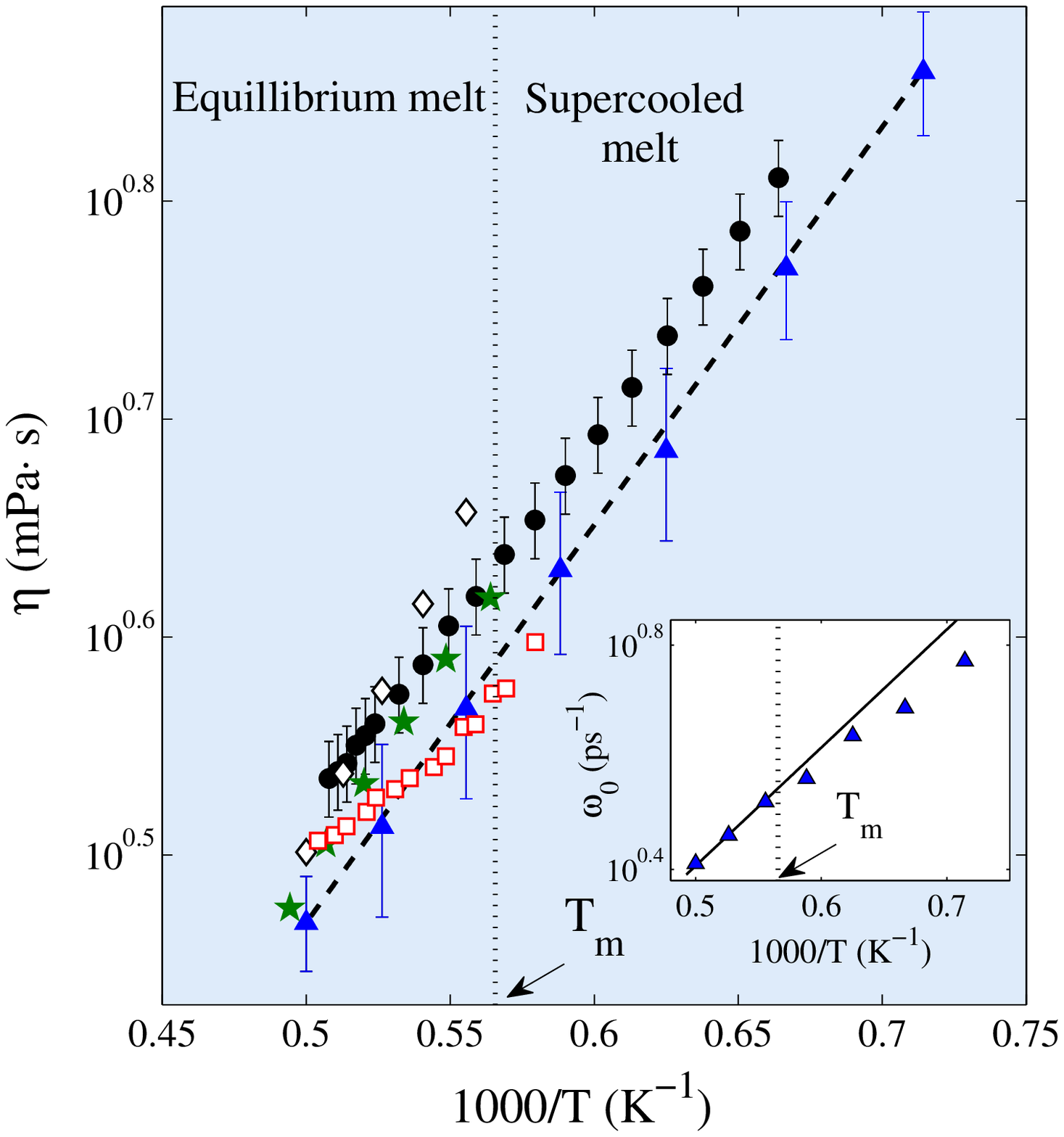}
\caption[kurzform]{\label{Fig_ShearVisc} (Color online)
\textbf{Main:} Temperature dependence of the shear viscosity for liquid cobalt at the pressure $p=1.5$~bar: markers ($\triangle~ \triangle~ \triangle$) represents the results of the molecular dynamics simulations calculated by Green-Kubo relation for the time correlation function of the stress tensor $S(t)$; ($\circ~ \circ~ \circ$) -- the experimental results; the dashed line represents the results of theoretical calculations performed according to the formula (\ref{Eq_KinVisc}); markers ($\diamondsuit~ \diamondsuit~ \diamondsuit$), ($\star~ \star~ \star$) and ($\Box~ \Box~ \Box$) are the experimental data taken from \cite{Assael2012}, \cite{HandBook} and \cite{Bodakin1978} respectively.
\textbf{Inset:} Time dependence of the characteristic frequency of atom vibration: markers ($\triangle~ \triangle~ \triangle$) represent the molecular dynamics data; The solid line -- the results of fitting by the Arrhenius law.}
\end{center}
\end{figure*}
%~~~~~~~~~~~~~~~~~~~~~~~~figure~~~~~~~~~~~~~~~~~~~~~~~~~~~~~~~~~~~~~~~~~~~

Fig. \ref{Fig_ShearVisc} shows the shear viscosity of liquid cobalt as a function of the inverse temperature in logarithmic scale. The plot is constructed on the basis of the experimental data, the simulation results, and the theoretical calculations with Eq.~(\ref{Eq_KinVisc}). As seen from Fig. \ref{Fig_ShearVisc}, the experimental data, simulation and theoretical results are reproduced fairly well by the Arrhenius law \cite{Frenkel}
\begin{equation}
\eta(T)=\eta_0\exp(E/k_BT). \label{Eq_Arren}
\end{equation}
Here, $\eta_0$ is the pre-exponential factor corresponding formally to the viscosity at $T\rightarrow\infty$; $E$ is the  activation energy. For our simulation results and the experimental data, we find the parameters $\eta_0^{MD}=0.29$~mPa$\cdot$s ($\eta_0^{Exp}=0.34$~mPa$\cdot$s) and $E^{MD}=6.6\cdot 10^{-20}$~J ($E^{Exp}=6.5\cdot 10^{-20}$~J), respectively.

Inset of Fig. \ref{Fig_ShearVisc} represents the temperature dependence of the characteristic frequency $\omega_0(T)$ of atomic vibrations found from Eq. (\ref{Eq_FreQ}). As can be seen, for the temperature range $T=[1800,2000]$~K, the dependence $\omega_0(T)$ is well reproduced by the Arrhenius law with the pre-exponential factor $\Omega_0=0.225$~ps$^{-1}$ and the activation energy $E=6.6\cdot 10^{-20}$~J, which coincides completely with the activation energy of the viscous process. Deviation from the Arrhenius law in the temperature dependence $\omega_0(T)$ below the melting temperature $T_m$ is due to the fact that Eq. (\ref{Eq_ViscM}) can be applied only to the equilibrium liquid.

Expression (\ref{Eq_KinVisc}) for the shear viscosity $\eta$ contains the frequency parameters $\Delta_1$ and $\Delta_2$, which are determined through the configuration characteristics of the system, namely, through the two and three particles distribution functions \cite{Tankeshwar}. Therefore, it is advisable to consider a possible correlation between the viscosity and structural characteristics of the system (for example, configuration entropy \cite{Abramson_2007,Abramson_2008,Abramson_2009}).

Probably, the most well-known expression, where such a relationship is assumed, is the Adam-Gibbs relation \cite{Adam/Gibbs}:
\begin{equation}
\eta=A\exp(C/TS_C).
\end{equation}
Here, $S_C$ is the configuration entropy and $C$ is the constant characterized the barrier height of atomic restructuring.
On the other hand, the Rosenfeld's scaling laws states relation between the excess entropy $S_{ex}$ and the transport coefficients such as the self-diffusion $D$, the viscosity $\eta$ and the thermal conductivity $\kappa$ of the system \cite{Rosenfeld_1977,Rosenfeld_1999}:
\begin{subequations}
\begin{equation}
D^*=D\frac{\rho^{1/3}}{(k_BT/m)^{1/2}}=A\exp{(-\alpha S_{ex})},
\end{equation}
\begin{equation}
\eta^*=\eta\frac{\rho^{-2/3}}{(mk_BT)^{1/2}}=B\exp{(\beta S_{ex})}, \label{Eq_Visc}
\end{equation}
\begin{equation}
\kappa^*=\kappa\frac{\rho^{-2/3}}{k_B(k_BT/m)^{1/2}}=C\exp{(\gamma S_{ex})}.
\end{equation}
\end{subequations}
Here, $A$, $B$, $C$, $\alpha$, $\beta$ and $\gamma$ are the property-specific constants which are equal for model fluids to $0.6$, $0.2$, $1.5$, $0.8$, $0.8$ and $0.5$, respectively \cite{Rosenfeld_1999}. The quantities $D^*$, $\eta^*$, $\kappa^*$ are dimensionless, the excess entropy $S_{ex}$ is given in units of $k_B$.
The Rosenfeld's scaling relation (\ref{Eq_Visc}) for the viscosity may be considered as an attempt to realize the following physical idea. Since the viscosity is simply proportional to the structural relaxation time, then according to relation (\ref{Eq_Visc}), the viscosity is proportional to the number of accessible configurations with the structural relaxation time \cite{Abramson_2009,Abramson_2011,Abramson_2014}.

The thermodynamic excess entropy is defined as the difference in entropy between the fluid and the corresponding ideal
gas under identical temperature and density conditions. The total entropy of a classical fluid can be written as \cite{Khusnutdinoff_2011}
\begin{equation}
S=S_{id}+\sum_{n=2}^NS_n,
\end{equation}
where $S_{id}$ is the entropy of the ideal gas reference state, $S_n$ is the entropy contribution due to $n$-particle spatial correlations. Then, the excess entropy is defined as
\begin{equation}
S_{ex}=S-S_{id}.
\end{equation}
The main contribution into $S_{ex}$ is due to the pair-correlation entropy $S_2$ which
is $\sim 85\div95 \%$ for the case of monatomic liquids over a fairly wide range of densities. Therefore, the next approximation can be applied: $S_{ex}\approx S_2$.
The pair-correlation entropy is defined as
\begin{eqnarray}
S_2=-2\pi \rho \int_0^{\infty}\bigg \{g(r)\ln(g(r))-[ g(r)-1]\bigg\} r^2 dr=\nonumber\\-2\pi \rho \int_0^{\infty}g(r)\ln(g(r)) r^2 dr+\frac{1}{2}\bigg(\rho k_BT\chi_T-1\bigg),
\label{Eq_S2}
\end{eqnarray}
where $\chi_T$ is the isothermal compressibility.

%~~~~~~~~~~~~~~~~~~~~~~~~figure~~~~~~~~~~~~~~~~~~~~~~~~~~~~~~~~~~~~~~~~~~~
\begin{figure*}
\begin{center}
\includegraphics[height=11cm, angle=0]{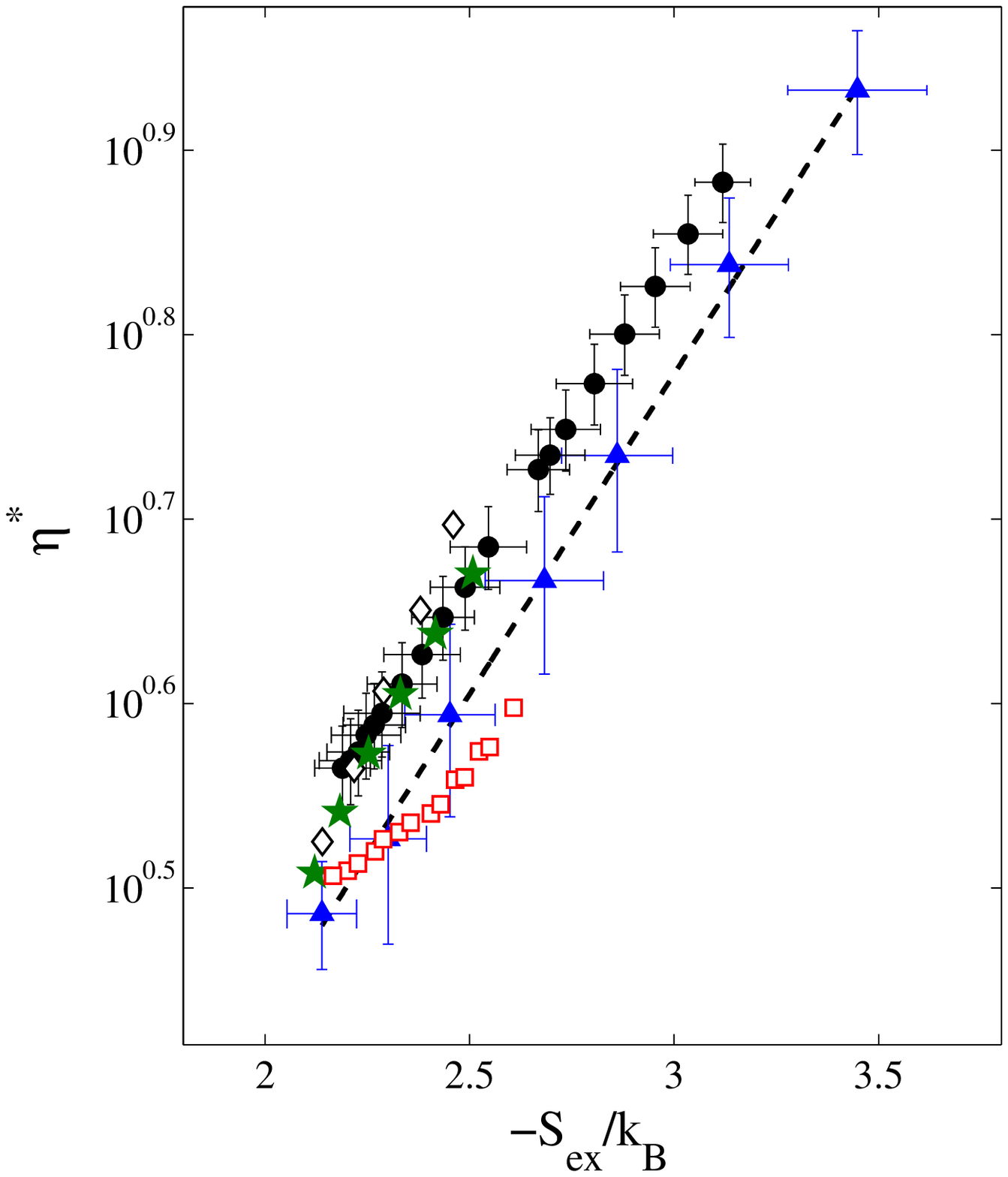}
\caption[kurzform]{\label{Fig_Visc} (Color online)
Rosenfeld's scaling law for the reduced shear viscosity $\eta^*$ with the pair-excess entropy $S_{ex}$ for liquid cobalt at the pressure $p=1.5$~bar: markers ($\triangle~ \triangle~ \triangle$) represent the results of molecular dynamics simulations calculated using the Green-Kubo relation for the time correlation function of the stress tensor $S(t)$; ($\circ~ \circ~ \circ$) -- the experimental results; solid line represents the results of the calculations performed according to the formula (\ref{Eq_KinVisc}); markers ($\diamondsuit~ \diamondsuit~ \diamondsuit$), ($\star~ \star~ \star$) and ($\Box~ \Box~ \Box$) are the experimental data taken from \cite{Assael2012}, \cite{HandBook} and \cite{Bodakin1978} respectively.}
\end{center}
\end{figure*}
%~~~~~~~~~~~~~~~~~~~~~~~~figure~~~~~~~~~~~~~~~~~~~~~~~~~~~~~~~~~~~~~~~~~~~

In Fig. \ref{Fig_Visc}, the reduced shear viscosity in logarithmic scale is shown at the corresponding values of the negative excess entropy $S_{ex}$ [using two-particle approximation (\ref{Eq_S2})]. First, our experimental values and experimental data from Ref. \cite{Assael2012,HandBook} as well as MD simulation results and theoretical results from Eq. (\ref{Eq_KinVisc}) yield the straight lines that is in agreement with the Rosenfeld's scaling representation (\ref{Eq_Visc}). Further, $\eta^*(S_{ex})$-dependence with experimental data from Refs. \cite{Assael2012,HandBook} differs from  these lines. Second, the Rosenfeld's scaling extends well to the temperature range of the supercooling melt, that is verified by the scaled representation of our experimental data as well as of our simulation and theoretical results. Third, our experimental data, simulation data and theoretical results for the shear viscosity yield the same slope in the Rosenfeld's scaling plot with the parameter
$\beta_{exp}=0.78\pm 0.02$ [see Eq. (\ref{Eq_Visc})]. Note that this value is close to $\beta\approx 0.8$, which is usually expected for the monoatomic fluids \cite{Rosenfeld_1999}. The prefactor $B$ is estimated to be $0.68$ and $0.6$ for our experimental data and theoretical (simulation) results for the viscosity, respectively; and $B=0.27$ for the data from Ref. \cite{Assael2012}.

\section*{\normalsize Conclusions}
In summary, the shear viscosity of liquid cobalt at different temperatures was determined  experimentally and numerically by means of molecular dynamics simulations with the EAM-interparticle interaction potential \cite{Passianot_1992}. Further, the spectral densities of the  stress tensor TCF $\tilde{S}(\omega)$ as well as the shear viscosity $\eta$ are computed within the framework of the microscopic theoretical model \cite{Mokshin_2003, Mokshin_2011}. It is found good agreement between theoretical results, experimental data and molecular dynamics simulations results for the viscosity of liquid cobalt. It is shown that experimental data as well as simulation and  theoretical results for the shear viscosity are reproduced by the Rosenfeld's model.

\section*{\normalsize Disclosure statement}
No potential conflict of interest was reported by the authors.

\section*{\normalsize Acknowledgments}
This work was supported by the Russian Foundation for Basic Research (project No. 18-02-00407-a). Authors are also grateful to the Ministry of Education and Science of the Russian Federation for supporting the research in the framework of the state assignment (No. 3.2166.2017/4.6).

\end{document}